# The Impact of the Revised Sunspot Record on Solar Irradiance Reconstructions

G. Kopp

Univ. of Colorado / LASP, Boulder, CO 80303, *USA*

Greg.Kopp@LASP.Colorado.edu

N. Krivova • C.J. Wu

Max-Planck-Institut für Sonnensystemforschung, Justus-von-Liebig-Weg 3, 37077 Göttingen, Germany

J. Lean

Space Science Division, Naval Research Laboratory, 4555 Overlook Avenue Southwest, Washington, DC 20375, USA

**Abstract** Reliable historical records of total solar irradiance (TSI) are needed for climate change attribution and research to assess the extent to which long-term variations in the Sun's radiant energy incident on the Earth may exacerbate (or mitigate) the more dominant warming in recent centuries due to increasing concentrations of greenhouse gases. We investigate potential impacts of the new Sunspot Index and Long-term Solar Observations (SILSO) sunspot-number time series on model reconstructions of TSI. In contemporary TSI records, variations on time scales longer than about a day are dominated by the opposing effects of sunspot darkening and facular brightening. These two surface magnetic features, retrieved either from direct observations or from solar activity proxies, are combined in TSI models to reproduce the current TSI observational record. Indices that manifest solar-surface magnetic activity, in particular the sunspot-number record, then enable the reconstruction of historical TSI. Revisions to the sunspot-number record therefore affect the magnitude and temporal structure of TSI variability on centennial time scales according to the model reconstruction methodologies. We estimate the effects of the new SILSO record on two widely used TSI reconstructions, namely the NRLTSI2 and the SATIRE models. We find that the SILSO record has little effect on either model after 1885 but leads to greater amplitude solar-cycle fluctuations in TSI reconstructions prior, suggesting many 18th and 19th century cycles could be similar in amplitude to those of the current Modern Maximum. TSI records based on the revised sunspot data do not suggest a significant change in Maunder Minimum TSI values, and comparing that era to the present we find only very small potential differences in estimated solar contributions to climate with this new sunspot record.

# 1. Introduction

Solar irradiance provides the energy that establishes Earth's climate. The next most significant energy sources are from radioactive decay and geothermal activity, although these combined with all other Earth heating sources are a factor of 2700 lower than the solar irradiance (Kren, 2015). Historically misnamed the "solar constant," the Sun's total radiant energy varies with time. Even small changes in this energy over long periods of time potentially affect Earth's climate, as demonstrated in modern times by Eddy (1976) and substantiated by more recent studies (Haigh, 2007; Lean and Rind, 2008; Ineson *et al.*, 2011; Ermolli *et al.*, 2013; Solanki, Krivova, and Haigh, 2013).

The spatially and spectrally integrated radiant energy from the Sun incident at the top of the Earth's atmosphere, which averages 1361 W m$^{-2}$, is now termed the total solar irradiance (TSI) and has been measured with space-borne instruments continuously since 1978. Overall increases of ≈ 0.1 % from the minimum to the maximum of the 11-year solar cycle are typical during recent decades (Fröhlich, 2006) with additional and occasionally larger variations occurring as sunspot and facular magnetic features emerge, transit, and decay on the solar disk facing the Earth. Figure 1 shows the observational record of TSI during the space era coincident with the sunspot number, illustrating that TSI fluctuations are in phase with activity levels over the solar cycle.

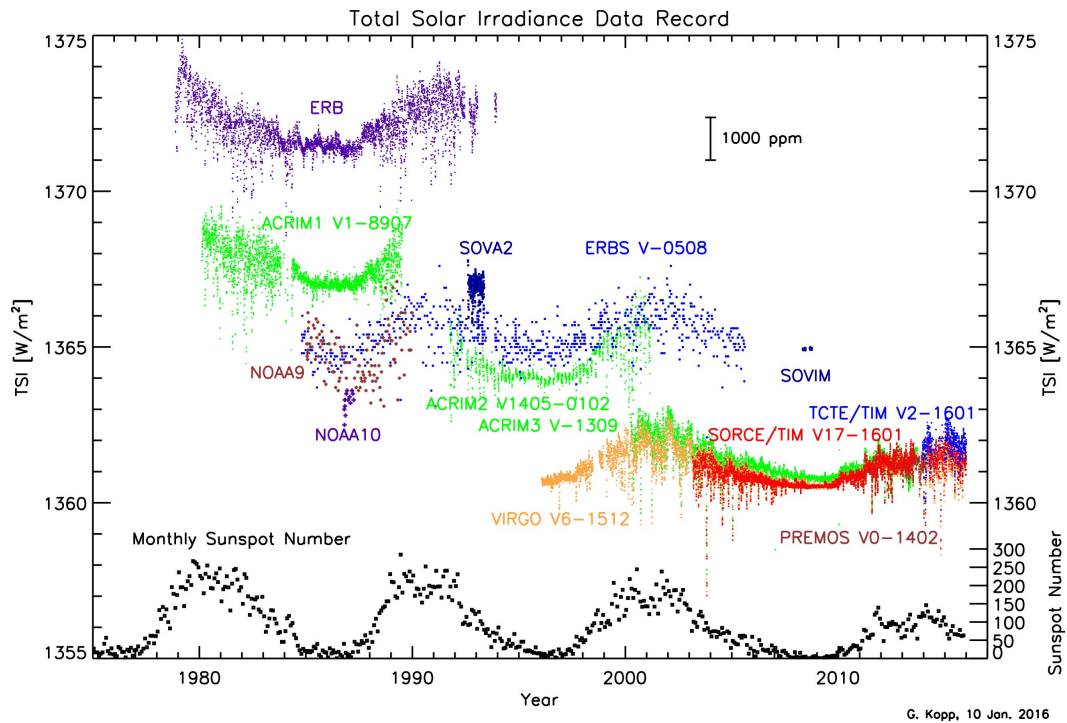

Figure 1: TSI measurements from more than a dozen space-borne instruments have been uninterrupted since beginning in 1978. These measurements show variations that are in phase with solar activity represented by the SILSO monthly sunspot number.

Evidence for the climate influence of solar cycle irradiance changes is apparent in surface and atmospheric temperatures (Lean, 2010; Gray *et al.*, 2010). A global surface-temperature increase of about 0.1 °C is associated with the irradiance increase in recent solar cycles, with larger regional changes occurring in some locations (Lean and Rind, 2008). In addition to its importance for climate change detection and attribution, reliable knowledge of potential secular changes in TSI is important for establishing climate sensitivity to natural influences, determined, for example, by extracting the climate response to purported historical solar irradiance variability from the observational surface-temperature record.

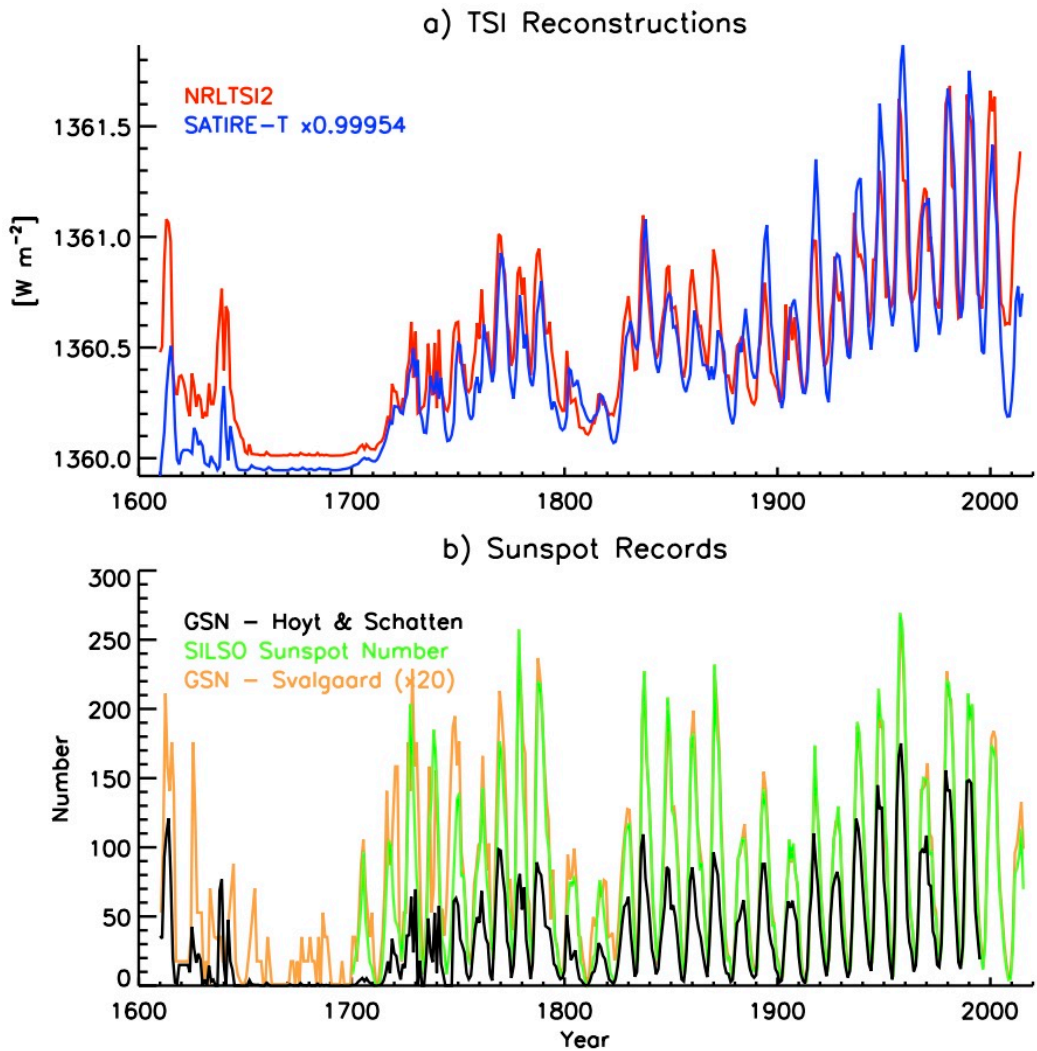

Figure 2: a) The current NRLTSI2 and SATIRE TSI reconstructions since 1610 are shown (upper plot). b) Prior to 1874 both models rely on the Hoyt and Schatten sunspot group number, which is compared to the new SILSO sunspot number and a revised group sunspot number by Svalgaard and Schatten (lower plot).

Solar irradiance variability can be modeled by correlating direct TSI observations with indices representative of solar-surface magnetic features, the primary ones being sunspots and faculae. Solar irradiance is then constructed at prior times by applying the correlations to historical indices of these solar features. A prominent "empirical" solar proxy model is the Naval Research Laboratory Total Solar Irradiance (NRLTSI: Lean, 2000; Lean *et al*., 2005; Coddington *et al*., 2015). To estimate solar irradiances extending back to 1610, the newest NRLTSI2 model (blue curve in Figure 2a) directly uses, in addition to sunspot area and facular indices after 1882, the sunspot-group-number time series of Hoyt, Schatten, and Nesmes-Ribes (1994) and Hoyt and Schatten (1998) developed from multiple historical sunspot observations, as shown in Figure 2b.

In the so-called "semi-empirical" models, solar magnetic-feature observations or proxies are applied less directly by assessing the solar disk area coverage from spots and faculae and then weighting the time-independent brightness contrasts of the corresponding components computed from solar model atmospheres. An example of a model of this type is the Spectral And Total Irradiance Reconstructions model (SATIRE), shown as the red curve in Figure 2a (Krivova, Vieira, and Solanki, 2010). Similarly to NRLTSI2, SATIRE relies on the historical sunspot record in the pre-spacecraft period. Details of and differences between these two models are described below.

Clette *et al*. (2015) have revised the sunspot-number record by reevaluating the > 400-year span of measurement records and normalizing for estimated offsets and biases among the numerous contributing observers. This method relies on a series of overlapping "backbones" based on the longest duration observations. Daisy-chaining these backbones and regressing them with shorter-term measurements for offset and slope normalizations are intended to achieve temporal continuity and to maintain consistency among various observers. The product of this effort is the new Sunspot Index and Long-term Solar Observations (SILSO) record.

As Figure 2b shows, the SILSO record differs notably from the Hoyt and Schatten (1998) sunspot group number used for many extant historical reconstructions. The method and the new record are discussed in more detail elsewhere in this issue (*e.g.* Usoskin *et al*., 2016; Lockwood *et al*. 2016). In this article we assess and discuss the impacts of these sunspot record differences on the NRLTSI2 and SATIRE solar irradiance reconstructions shown in Figure 2a. While there are other solar variability models that extend the TSI record prior to 1610 by using cosmogenic isotopes (Vieira *et al*., 2011; Delaygue and Bard, 2011; Steinhilber *et al*., 2012; Usoskin, 2013), the current article evaluates the effects of the recently updated sunspot record on TSI reconstructions from 1610 to the present, which includes the time domains of the SILSO record and those of the NRLTSI2 and SATIRE models that directly use the sunspot record during this period.

In this article we summarize the NRLTSI2 and SATIRE model formulations and estimate the changes to their TSI reconstructions that result from differences between the sunspot group number records currently used in the models and the SILSO sunspot-number record. We then consider how the correspondingly revised TSI reconstructions may affect the determination of climate sensitivity to solar forcing and thus the climate change attributable to solar variability in recent centuries.

## 2. NRLTSI2 Model

### 2.1 Model Formulation

The NRLTSI2 model (Coddington *et al*., 2015) calculates total solar irradiance [$T(t)$] at a specified time [$t$] assuming that bright faculae and dark sunspots on the solar disk alter the baseline "quiet" total solar irradiance [$T_Q$] by amounts $\Delta T_F(t)$ and $\Delta T_S(t)$, respectively, so that $T(t) = T_Q + \Delta T_F(t) + \Delta T_S(t)$. The model is part of the new NOAA Solar Irradiance Climate Data Record, the construction, development, transition, validation, and operation of which Coddington and Lean (2015) describe in detail.

NRLTSI2 is a new version of the NRLTSI model developed initially by Lean (2000) using a composite record of TSI constructed from multiple observations made between 1978 and 1998 (Fröhlich and Lean, 1998). In addition to contemporary sunspot and facular sources, the initial version of NRLTSI included a long-term component of TSI variability derived from variations in Sun-like stars (Lean, Skumanich, and White, 1992). In a subsequent revision of the model, Lean *et al.* (2005), motivated by uncertainty about the veracity of the Sun-like star observations (Radick *et al.*, 1998), reduced the magnitude of the assumed background component by a factor of about four to reflect estimates of magnetic-flux changes on the Sun since 1710 made using a flux transport model (Wang, Lean, and Sheeley, 2005).

In NRLTSI2, $T_Q$ = 1360.45 W m$^{-2}$ based on measurements of total solar irradiance made by TIM on SORCE (Rottman, 2005; Kopp and Lawrence, 2005; Kopp and Lean, 2011) during solar minimum conditions when the solar disk is mostly free of faculae and sunspots. The approach for determining the two time-varying model components [$\Delta T_F(t)$] and [$\Delta T_S(t)$] is to assume that each relates linearly to an appropriate facular [$F(t)$] and sunspot [$S(t)$] index. Thus, the TSI model [$T_{\text{mod}}$] is formulated from TSI observations [$T_{\text{obs}}$] as

$$T_{\text{mod}}(t) = T_{\text{obs}}(t) - T_Q = a_0 + a_1 F(t) + a_2 S(t) + \varepsilon$$

where the model coefficients $a_0$, $a_1$, and $a_2$ are estimated from multiple regression using SORCE TIM V.17 TSI observations made between 2003 and 2014, and $\varepsilon$ is the minimized residual error. The indices for the facular brightening [$F(t)$] and sunspot darkening [$S(t)$] are obtained from ground- and/or space-based observations.

In the extant spacecraft era, the proxy of facular brightening [$F(t)$] is the University of Bremen composite record of measurements of the irradiance (*i.e.* integrated over the solar disk) of Magnesium (Mg) II emission. A Mg II index is determined as the ratio of measurements from the core of the H and K Mg II emission lines at 280 nm to measurements in the nearby wings (278 and 282 nm). The variability in this index is attributed to chromospheric extensions of the photospheric faculae. Since 1978, multiple spacecraft missions have recorded the Mg II index (Skupin *et al.*, 2004; Viereck *et al.*, 2004; Snow *et al.*, 2014). The proxy of sunspot darkening [$S(t)$] is computed from the areas and locations of sunspots on the solar disk on any given day (Allen, 1976; Foukal, 1981; Brandt, Stix, and Weinhardt, 1994; and Lean *et al.*, 1998) as reported by the Air Force Solar Observing Optical Network (SOON) sites. Royal Greenwich Observatory (RGO) observations provide sunspot-region information prior to 1982, albeit with some uncertainty associated with establishing their cross calibration with the SOON observations (Foukal, 2014).

The coefficients of the NRLTSI2 model are explicitly determined using TIM TSI observations, the University of Bremen Mg II facular index, and sunspot darkening calculated from the SOON network for the period 2003 to 2014, which is the duration of the TIM observations. Figure 3 compares the NRLTSI2 model with direct TIM observations and delineates the separate sunspot and facular components that produce TSI variability in the spacecraft observation era (since 1978). The correlation of the model and the observations is 0.96, the standard deviation of the residuals is 0.1 W m$^{-2}$, and the slope of the residuals is 1 ppm per year, which is less than the TIM's estimated 10 ppm per year stability uncertainty. Time-dependent uncertainties in the model's irradiance variability are estimated by combining uncertainties in the input facular-brightening and sunspot-darkening indices (typically ± 20 %) with statistical uncertainties in the model coefficients, taking into account autocorrelation in the residuals. Uncertainties are larger prior to the spacecraft era because of the need to cross calibrate extant historical solar-activity indices with those used for the model formulation and because of the lack of knowledge of longer-term (multi-decadal) irradiance changes.

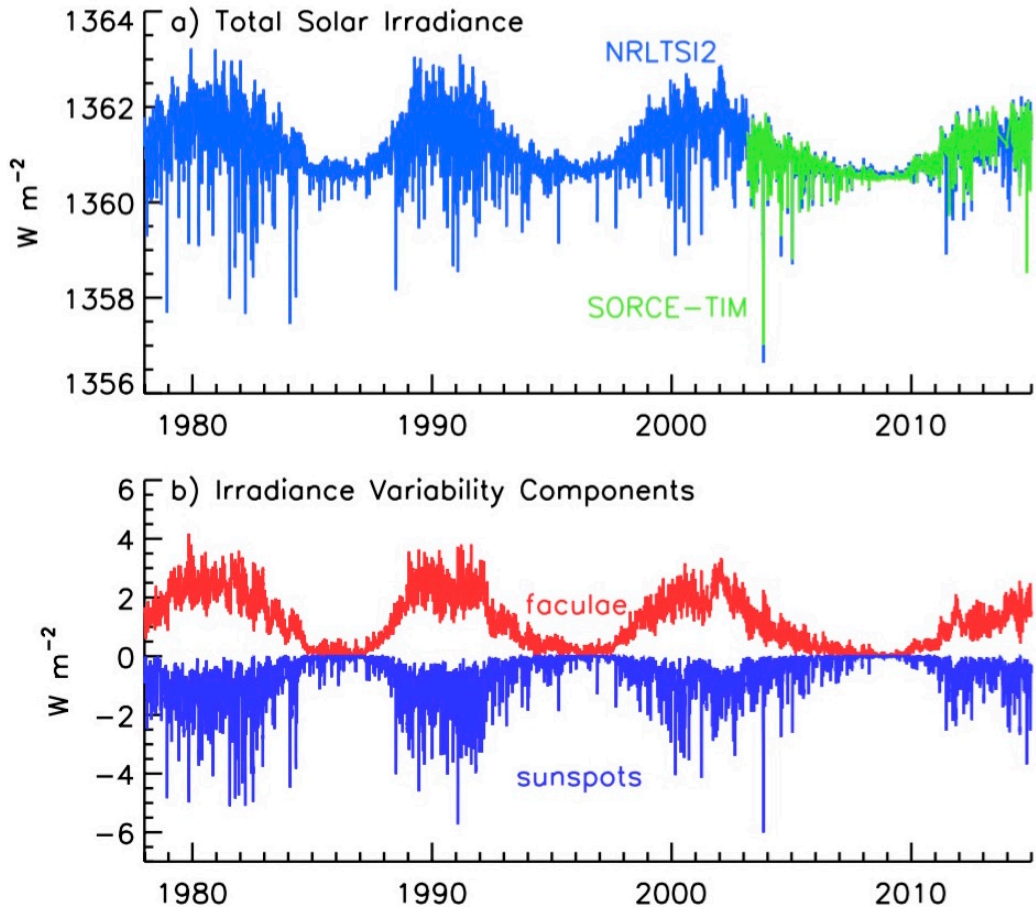

Figure 3: The NRLTSI2 model shows excellent agreement with SORCE TIM observations (upper plot). A decomposition into the opposing sunspot and facular components indicates the contributions of each to the model (lower plot).

Estimates of TSI, such as shown in Figure 2a from the 17th century Maunder Minimum period of anomalously low solar activity to the contemporary space-based TIM observations, are calculated by inputting to the model a combination of cross-calibrated sunspot and facular indices available in different epochs. While the sunspot and facular indices estimate solar cycle changes, an additional long-term facular contribution produces secular irradiance changes speculated to underlie the solar activity cycle at times prior to 1950. According to simulations of eruption, transport, and accumulation of magnetic flux on the Sun's surface since 1617 using NRL's magnetic flux transport model including variable meridional flow, a small accumulation of total magnetic flux and possibly the rate of emergence of small, magnetic bipolar ("ephemeral") regions on the quiet Sun can produce a net increase in facular brightness (Wang, Lean, and Sheeley, 2005). The resulting modeled increase in TSI from the Maunder Minimum to the present-day quiet Sun is about 0.04 % (see estimates by Lean *et al.*, 2005). Since this background component is speculative, the associated uncertainty in the reconstructed TSI on these time scales is equal to the magnitude of the adopted background component itself.

## 2.2 Temporal Regions of Input Indices

There are several distinct temporal regions in which the NRLTSI2 reconstruction utilizes different solar activity indices. The sunspot-number record is used only prior to 1950.

i) During the time period of space-borne irradiance observations (1978 to the present), the NRLTSI2 facular index is the University of Bremen Mg II core-to-wing ratio and the sunspot darkening index is calculated from sunspot areas and locations from SOON observations archived at the NOAA World Data Center.

ii) Prior to 1978 the Mg index does not exist so, as Lean *et al.* (2001) describe in detail, an equivalent facular index is constructed. This index is based on Ca K plage observations, which provide information about facular changes arising from the largest active regions, and on the smoothed 10.7-cm radio flux [$F_{10.7}$] which represents additional solar-cycle variability due to facular emission from accumulated smaller regions in the enhanced network. These indices extend back to 1950, which is when the $F_{10.7}$ radio flux observations commenced.

iii) Prior to 1976 the sunspot areas and locations are those measured by the RGO from 1882 to 1976 (Hoyt and Eddy, 1982). RGO observations are recognized as reporting larger sunspot areas than those measured by SOON, but the exact scaling is uncertain. Hathaway, Wilson, and Reichmann (2002) suggest that RGO areas are 40 % to 50 % larger than the SOON areas, consistent with Balmaceda *et al.* (2009, 2010), whereas Foukal (2014) determines a 20 % higher value for the RGO areas. Fligge and Solanki (1997) also provide additional insight into this scaling from a cross comparison using Rome observations. The extant NRLTSI2 model scales RGO areas to 80 % of their given values to match the spacecraft-era SOON data (*i.e.* the RGO areas are assumed to be 25 % larger than the SOON areas, consistent with Foukal). To indicate the model's sensitivity to scaling from the multiple different published results, Figure 5 compares this NRLTSI2 reconstruction with one based on a 67 % scaling factor (*i.e.* the RGO areas are assumed to be 49 % larger than the SOON areas). This latter value gives a composite sunspot blocking record that has the highest correlation with the sunspot number and is in good agreement with the scaling provided by Balmaceda *et al.* (2009) (*cf.* also Fligge and Solanki, 1997). The differences in TSI reconstructions due to selection of RGO sunspot area scaling factor are small, however, because of the significant relative contributions to the reconstructions from the facular index, which is independent of RGO areas.

iv) Prior to 1950 the sunspot number is used to estimate the solar-cycle variations in the facular index, extending this index back to 1882 and enabling reconstructions prior to the commencement of 10.7-cm flux observations. The sunspot-darkening index is obtained separately by using the (scaled) RGO sunspot area observations.

v) Prior to 1882 the RGO sunspot observations are not available, precluding the determination of separate facular and sunspot indices for inputs to the model. Lacking these inputs, the model utilizes the sunspot number as its sole proxy to estimate the net solar-cycle changes in irradiance. Since sunspot and facular components cannot be separated in this early period because there is only the one sunspot-number index available, the TSI is calculated only annually.

An additional background component based on NRL flux transport model simulations by Wang, Lean, and Sheeley (2005) is added to the solar-cycle variations determined from the sunspot and facular components to estimate secular variations when extending the reconstructions back in time. That flux transport model itself uses the sunspot-number record, so the background component tracks the smoothed sunspot number. In this article we do not recalculate the extensive flux transport simulations, but rather illustrate the expected scenario for the SILSO sunspot-number record with a background component consistent with the flux transport simulations.

## 3. SATIRE Model

### 3.1 Model Formulation

In the SATIRE model (Solanki and Krivova, 2006; Krivova, Solanki, and Unruh, 2011), irradiance variations are attributed to changes in the surface distribution and coverage by photospheric magnetic components, namely sunspot umbrae and penumbrae, faculae, and the network. The surface free of such detectable features is considered to be the quiet Sun. Each component is ascribed a typical, time-invariant brightness spectrum calculated from the corresponding semi-empirical solar model atmospheres (Unruh, Solanki, and Fligge, 1999). These brightness spectra are weighted by their corresponding disc area coverage on each day derived from solar-activity proxies and summed to give the brightness spectrum of the whole Sun. The modeled TSI is an integral over the entire resulting spectrum.

### 3.2 Temporal Regions of Input Indices

SATIRE irradiance variations are derived from changes in the amount and the distribution of the solar-surface magnetic field, either observed directly or reconstructed from appropriate proxies using a physical model. As with the NRLTSI2 model, there are distinct temporal regions in which the reconstruction utilizes different proxy indices, with many being shared between the two models particularly at earlier epochs when fewer solar observational records are available.

i) From 1974 to the present the SATIRE model uses direct measurements of the solar surface magnetic field. These are based on magnetograms, many of which come from space-borne instruments (Yeo *et al*., 2014). The specific model in the spacecraft era period is referred to as the "SATIRE-S" model and overlaps with multiple direct measurements of the TSI. The correlation coefficient between the model and the PMOD observational composite is 0.96 over the period from 1978 to 2014 and 0.98 for the period after 1996 when space-based magnetograms (which are not degraded by ground-based atmospheric seeing) are available.

ii) During the period from 1874 to 1974, SATIRE primarily uses the sunspot area and position record, which is based on the RGO, Russian, and SOON observations as described by Balmaceda *et al*. (2009), but also partly on the sunspot number, with the default being the group sunspot number from Hoyt and Schatten (1998). There are two versions of the model:

    a) In SATIRE-T (Krivova, Balmaceda, and Solanki, 2007; Krivova, Vieira, and Solanki, 2010), the darkening due to sunspots is based on the sunspot-area record rather than a mere count of sunspot number, and thus is unaffected by the choice of the sunspot-number record. To retrieve the evolution of the bright components (faculae and the network), a simple physical model described by Solanki, Schüssler, and Fligge (2002) is used. The subsequent evolution of the surface magnetic field is described by a set of ordinary differential equations, whereby the sunspot number serves as a proxy for the emergence rate of the magnetic flux in active and ephemeral regions. As spot contribution is known from the spot area observations, the sunspot record used only affects the bright components of the model.

    b) SATIRE-T2 (Dasi-Espuig *et al*., 2014) uses the surface-flux transport model by Jiang *et al*. (2010) to describe the evolution of the active regions (including both spots and faculae) and the open magnetic flux. The evolution of the weaker ephemeral regions forming the

network is calculated using the sunspot number as input. The amplitude of the variation of this component is small compared to the changes in coverage of active regions.

iii) Similarly to the NRLTSI2 model, from 1610 to 1874 the sunspot number is the only input into SATIRE-T. The reconstruction discussed here and shown in Figure 2 is based on the GSN from Hoyt and Schatten (1998). The choice of sunspot-number record during these times thus influences the model directly.

## 4. Impact of SILSO Sunspot Number on TSI Reconstructions

Since the NRLTSI2 and SATIRE reconstructions in the spacecraft era are based on a variety of cross-calibrated solar indices other than the sunspot number, the SILSO sunspot record primarily affects earlier historical epochs. The subsections below give the specific effects on each TSI model.

The estimates presented here are initial results. More detailed future calculations are needed to fully assess the differences caused by the new SILSO record. For NRLTSI2 a more complete assessment would require a recalculation of the evolution of total and closed magnetic flux in response to changing solar cycle amplitudes, as done by Wang, Lean, and Sheeley (2005).

### 4.1 SILSO Effects on NRLTSI2

Figure 4 quantitatively compares the new SILSO record with the Hoyt and Schatten group sunspot number. From 1885 to 2014 the two are highly correlated (correlation coefficient > 0.99) and a linear regression reliably transforms one to the other *via* a simple scaling factor. Figure 4b shows the good agreement of the scaled Hoyt and Schatten group sunspot number and the SILSO sunspot record after 1885. However, the relationship differs for values from 1700 to 1884. Although still relatively well correlated (correlation coefficient > 0.88 over this earlier era), the new sunspot record has larger values relative to the group sunspot number scaled according to the values after 1885.

The excellent overlap of the group sunspot numbers (used in the NRLTSI2 model) scaled to the SILSO sunspot-number record means that the SILSO record will give similar reconstructions as the currently used record, and therefore has minimal impact on the NRLTSI2 values after 1885. The primary effect of the SILSO record is on TSI reconstructions prior to 1885. SILSO-based estimates of annual TSI at these times are determined by correlating annual averages of SILSO-based NRLTSI2 values after 1885 with the similarly averaged sunspot record at those times and then applying that correlation to the SILSO record from 1700 to 1885. Figure 5 compares annual TSI variations modeled this way using the SILSO record with the extant values for the NRLTSI2 model for both 80 % and 67 % scalings of RGO sunspot areas; the differences due to these scalings, as stated in §2, are small compared to the overall differences between the NRLTSI2 reconstruction based on the Hoyt and Schatten group sunspot number and the new SILSO record.

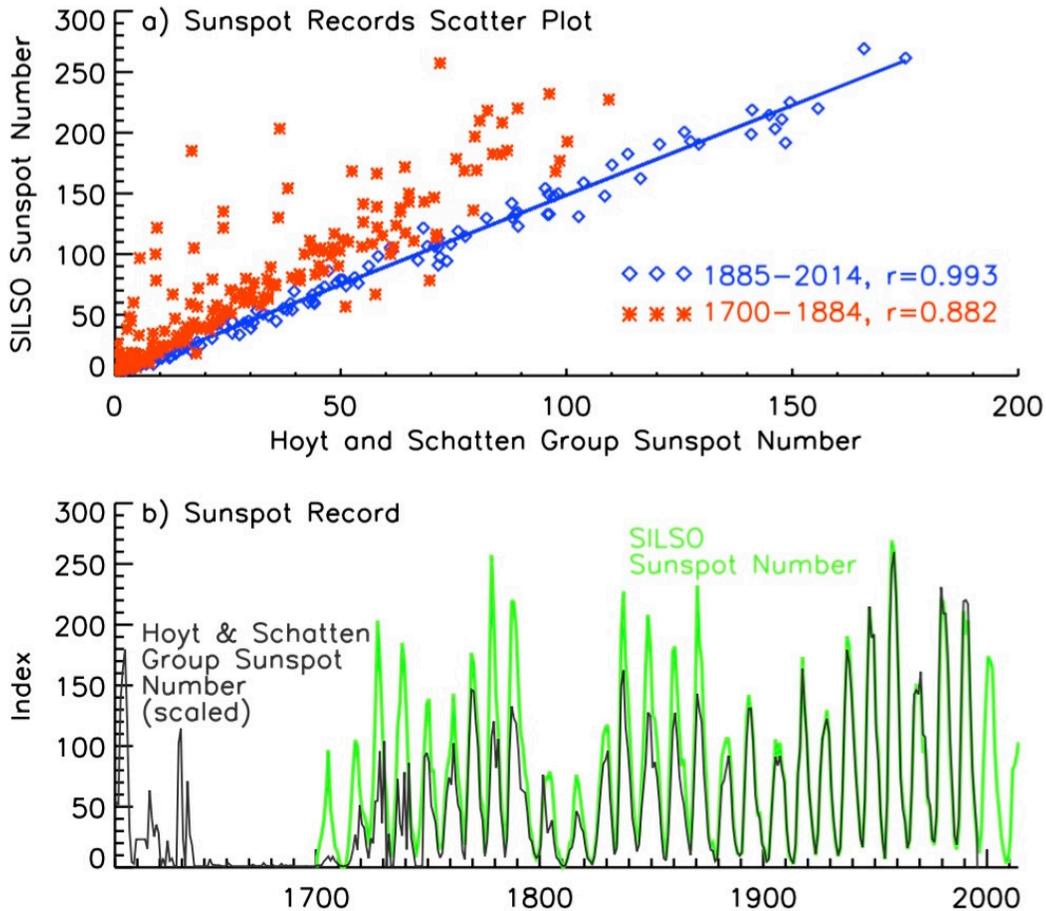

Figure 4: The Hoyt and Schatten group sunspot number scales with the SILSO values from 1885 to the present, although earlier times have less linear agreement (upper plot). A time series of the Hoyt and Schatten values scaled to match the SILSO record from 1885 to the present is compared with the SILSO record (lower plot).

One primary effect the SILSO record has on the NRLTSI2 historical TSI reconstruction is that most solar cycles exhibit larger amplitudes than in the current reconstruction during earlier eras. From 1700 to 1885, where the model is most affected by choice of sunspot record, the SILSO-based NRLTSI2 reconstruction has 0.018 % greater average TSI and solar cycle amplitudes averaging 1.5 times larger than in the current NRLTSI2 reconstruction.

The SILSO-based NRLTSI2 model values plotted in Figure 5 that extend through the 17th century are scaled from the existing NRLTSI2 model during that period, with that scaling factor determined from ratios between the two reconstructions during the first few decades after 1700. The SILSO-based reconstruction suggests that TSI may be slightly higher during the Maunder Minimum than does the existing NRLTSI2 reconstruction. With the extant NRLTSI2 reconstruction, the irradiances during the 17th century Maunder Minimum are 1360.0 W m$^{-2}$, or 0.033 % lower than the contemporary quiet-Sun value of $T_Q$ = 1360.45 W m$^{-2}$. Using the existing NRLTSI2 model values scaled to the SILSO-based reconstruction, this reduction is 0.023 % of the current quiet-Sun value, giving a value of 1360.14 W m$^{-2}$ and suggesting that the Maunder Minimum may have had slightly higher average irradiances than previously modeled.

The Impact of the Revised Sunspot Record on Solar Irradiance Reconstructions, Kopp *et al.*, 2016     p. 10

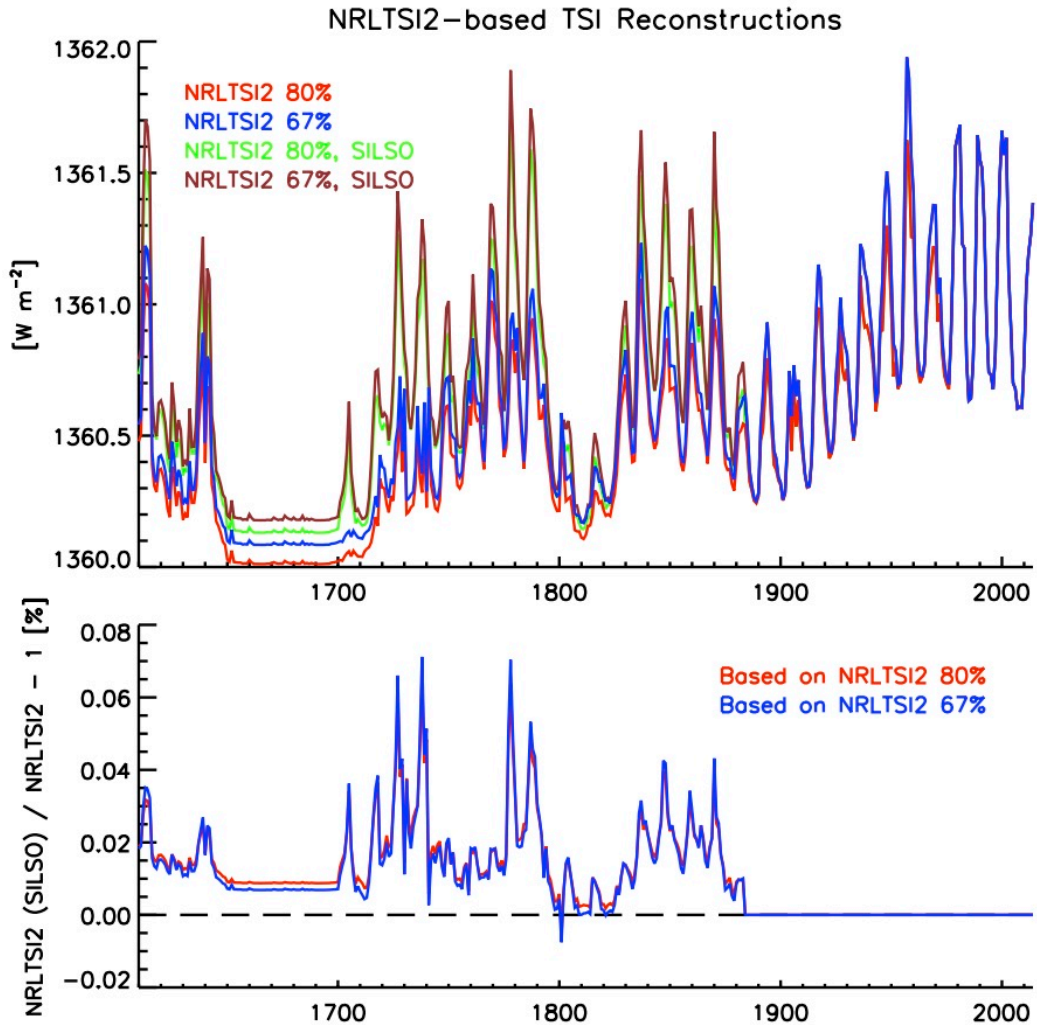

Figure 5: (upper plot) A NRLTSI2 reconstruction based on the new SILSO record (green) shows larger solar cycle amplitudes prior to 1885 and a slightly higher Maunder Minimum value than that based on the Hoyt and Schatten sunspot group number (red) from the extant NRLTSI2 model using a 80 % scaling of RGO sunspot areas. For comparison, reconstructions using a 67 % scaling of those areas are also shown (blue for the NRLTSI2 model and brown for the SILSO-based reconstruction from that model). The ratio of the SILSO-based reconstructions to the NRLTSI2 models (using both scalings) are shown in the lower plot and are generally <0.05 % of the TSI level. (Since the SILSO sunspot number used in these reconstructions does not extend prior to 1700, earlier dates in these plots are merely from the existing NRLTSI2 reconstruction scaled to the SILSO-based values averaged over a few decades after 1700.)

## 4.2 SILSO Effects on SATIRE

The SATIRE-S model, covering 1974 to the present, uses magnetograms to provide the spatial locations and intensities of magnetic features on the solar disk. In this era the model does not rely on the sunspot record and is therefore completely unchanged by the new SILSO record.

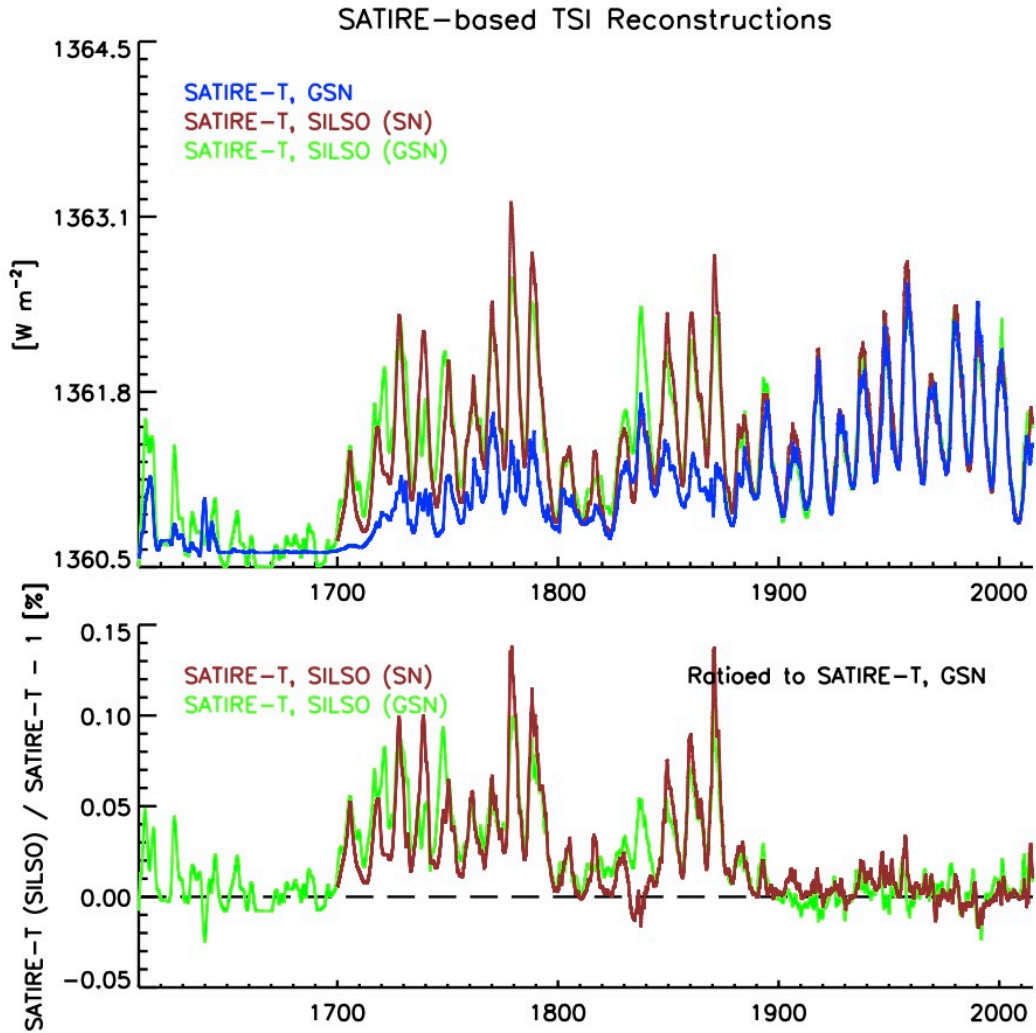

Figure 6: SATIRE-T reconstructions are compared for different proxy inputs indicating solar surface magnetic features (upper plot). During the spacecraft era, SATIRE-T does not rely on the sunspot number, and is therefore unaffected by the SILSO record. Comparisons of the current SATIRE-T (blue), based on group sunspot number (GSN), with a SATIRE-T reconstruction based on the SILSO sunspot-number record (brown) and on the Svalgaard and Schatten GSN (green) show differences in pre-spacecraft era times. The ratios of these reconstructions to that of SATIRE-T are shown in the lower plot, showing differences are generally <0.1 % of the TSI level. From 1874 to 1974 the effects of using the SILSO record on SATIRE results are minor, as the model has only an indirect reliance on sunspot number during this era. Before 1874, however, the SILSO sunspot record leads to greater estimates of solar activity than the current SATIRE reconstructions, indicating possibly higher overall irradiances and larger solar cycle amplitudes than current estimates in both the 18th and 19th centuries but similar mean TSI levels during Maunder Minimum.

During 1874 to 1974, although the sunspot component of the SATIRE-T model does not rely on (and is therefore unaffected by choice of) the sunspot number, the bright components are estimated from a physical model that uses the sunspot number as a proxy for the emergence rate of magnetic flux. Thus SATIRE-T can be affected by the choice of sunspot record in this era. However, owing to the comparatively small differences between the various sunspot-number records in this time frame, the resulting TSI reconstructions using different sunspot-number records are quite similar. SATIRE-T2, which uses the sunspot number only to estimate the effects of ephemeral regions in this era, is even less directly affected by the SILSO record. Figure 6 compares the extant SATIRE-T reconstruction with that based on the SILSO record. The SILSO-based SATIRE reconstruction has < 0.007 % greater average TSI and solar cycle amplitudes averaging 1.16 times larger during the 1874 – 1974 era. The effects of different sunspot-number records on the SATIRE-T2 irradiance reconstructions are nearly insignificant during this time because ephemeral regions have only a small effect on the amplitude of the TSI compared to sunspots and bright regions, and the differences among various sunspot-number records used to estimate the ephemeral regions are very small over this period.

Prior to 1874, since SATIRE relies primarily on the sunspot number going back to 1610, the effects of the new SILSO record are more direct. The significantly higher solar activity indicated by the SILSO sunspot record results in a correspondingly 0.03 % higher overall level of SATIRE TSI reconstructions in the 18th and 19th centuries, as shown in Figure 6. Average solar-cycle amplitudes are a factor of 3.2 larger in the SATIRE-T TSI reconstruction based on the SILSO record during these times than for the existing model. The SILSO sunspot-number record has a much greater effect prior to 1874 on the SATIRE model than on the NRLTSI2 model due to the different models' sensitivities to and use of the sunspot-number record. According to this SATIRE reconstruction, the Sun was brighter over extended periods in this era than during the last 3.5 solar cycles covered by direct spacecraft measurements, with some historical peaks comparable in magnitude to the large variation during Solar Cycle 19 (peaking in the late 1950's). Similarly, the Dalton Minimum (1796 to 1830) and the early 20th century minima are more pronounced in reconstructed TSI when using the SILSO record. The average TSI at activity-cycle maxima at these times could thus be comparable to the average TSI during activity minima of many other solar cycles. We note, however, that according to long-term TSI reconstructions based on cosmogenic-isotope data (Steinhilber, Beer, and Fröhlich, 2009; Delaygue and Bard, 2011; Vieira *et al.*, 2011), which are independent of sunspot observations, the TSI level in the 17th and 18th centuries remained clearly below that of the Modern Maximum.

# 5. Impact of the Revised Group Sunspot Number on TSI Reconstructions

The SILSO sunspot number does not extend prior to 1700, but a revised group sunspot number (GSN) record extending to 1610 has been constructed by Svalgaard and Schatten (2015) and is now available as well. Although based on many of the same observational records, the GSN only counts sunspot groups.

In addition to the TSI reconstruction based on the SILSO sunspot-number record described in §4, Figure 6 includes a SATIRE-T reconstruction extending back to 1610 based on the Svalgaard and Schatten GSN record. Other than the extension to earlier times, differences between the two reconstructions are small with the exception of Solar Cycle 8 peaking in the late 1830's. We also find differences between the SATIRE reconstruction based on this revised GSN and the existing SATIRE-T model to be an insignificant 0.001 % in terms of mean absolute value through the Maunder Minimum, although the reconstruction based on the GSN from Svalgaard and Schatten (2015) indicates more variability during that extended period of low solar activity. Even more so than the NRLTSI2 model based on the SILSO sunspot-number record extended to these earlier times, this SATIRE reconstruction suggests that the overall TSI level during the Maunder

Minimum would be much as currently modeled but that the revised GSN-based TSI reconstruction would have risen at a faster rate and to higher levels coming out of that period.

## 6. Implications for Understanding Climate Sensitivity and Climate Change

A change in radiative climate forcing [$\Delta F$] whether by varying solar irradiance or greenhouse gas concentrations, alters Earth's energy balance, requiring that surface temperatures seek a new equilibrium value. The global surface temperature change [$\Delta T$] arising from a forcing depends on the climate sensitivity [$\kappa$] such that $\Delta T = \kappa \Delta F$, where the climate sensitivity encompasses feedback processes involving primarily clouds, water vapor, and ice to the forcing (*e.g.* Hartmann, 1994). Climate sensitivity is uncertain by a factor of about two and may differ for different forcings but is generally considered to be in the range 0.2 to 1 $^o$C per W m$^{-2}$ forcing. Also uncertain is the response time for the climate system to reach equilibrium, which depends primarily on the amount of energy transported through the near-surface ocean mixed layer to the deep ocean.

An increase in TSI of 1 W m$^{-2}$ produces solar forcing $\Delta F_{sol} = 0.7 \times 1/4 = 0.18$ W m$^{-2}$, where the scalar factors account for the Earth's albedo and the geometric illumination of the entire Earth's surface by the Sun. Assuming a climate sensitivity to solar forcing of 0.6 $^o$C per W m$^{-2}$ (*i.e.* midway in the estimated range), an increase in TSI of about 1 W m$^{-2}$ from the 17th century Maunder Minimum to the present solar-cycle average produces global warming of 0.1 $^o$C. This estimated warming is reduced about 20 % to 0.08 $^o$C via the NRLTSI2 model and not reduced at all via the SATIRE model when using the new sunspot record to reconstruct TSI. The difference is small and within the uncertainty of the TSI reconstructions themselves (Figure 5 and Figure 6). Furthermore, this possible solar-caused global surface-temperature increase is significantly lower than the net measured temperature increase of at least 0.8 $^o$C over the same four-century time frame.

Although the net difference in forcing between the Maunder Minimum and the present may not be significantly different due to the SILSO record, as indicated in §4 and §5, the SILSO-based TSI reconstructions do suggest a different temporal variability of solar forcing. With the SILSO-based TSI solar cycle amplitudes from the early 1700's to the late 1800's being 1.5 to 3.2 times larger than those of the current reconstructions, the average TSI increase since the early 1700's would have been more gradual than in runs based on the traditional sunspot number record and the increase in TSI at the end of the Maunder Minimum would have been more abrupt than current reconstructions suggest. Climate model runs that use SILSO-based TSI reconstructions are needed to evaluate the resulting effects of this new sunspot number over the last three centuries.

## 7. Conclusions

From 1885 to the present, the effect of the SILSO record on both the NRLTSI2 and the SATIRE-T TSI reconstruction models is nearly insignificant for two reasons: the models' reliance on the sunspot record is small; and the differences between the SILSO and extant sunspot records are themselves small.

The revised sunspot record primarily causes changes in the historical TSI reconstructions between 1700 and 1874, when the sunspot record is the sole proxy for both the NRLTSI2 and the SATIRE models. The reconstructed TSI solar-cycle amplitudes in this time frame are approximately a factor of 1.5 times larger than the current NRLTSI2 model and 3.2 times larger than the current SATIRE-T model, giving overall solar irradiances that are about 0.02 % to 0.035 % higher and being comparable to those of the Modern Maximum at the end of the 1900's. This would suggest lower overall solar forcing during the last three centuries than from current estimates. Reconstructions or extrapolations back to the Maunder Minimum suggest that the mean solar irradiance at these times of low solar activity is nearly identical whether using the current model reconstructions or those based on the SILSO record.

The net effect on climate due to this new sunspot-number record is at a nearly insignificant level. Using the SILSO record, global warming estimates attributable to solar variability over the last four centuries may be up to 20 % less than current estimates suggest, perhaps causing only a 0.08 °C increase rather than a possible 0.1 °C increase in global surface temperature. These differences are negligible compared to uncertainties, and either contribution to overall global warming remains much less than those due to other climate influences.

**Acknowledgements**

We gratefully acknowledge the support of NASA's SORCE (NAS5-97045) and SIST (NNX15AI51G) for this effort. J. Lean appreciates collaboration with Odele Coddington in developing the NOAA Solar Irradiance CDR. The authors also appreciate helpful suggestions from the article's reviewer. The SILSO data are courtesy of WDC-SILSO, Royal Observatory of Belgium, Brussels. Figure 1 includes data from: www.ngdc.noaa.gov/stp/SOLAR/solar.html (NIMBUS7/ERB, ERBS/ERBE, NOAA9, and NOAA10); http://www.acrim.com (ACRIM1, ACRIM2, and ACRIM3); the VIRGO team via ftp://ftp.pmodwrc.ch; http://lasp.colorado.edu/home/sorce/data/tsi-data/ (SORCE/TIM); the PICARD/PREMOS team (personal communication); and http://lasp.colorado.edu/home/tcte/data/ (TCTE/TIM).

**Disclosure of Potential Conflicts of Interest**

The authors declare that they have no conflicts of interest**.**